\documentclass[acmsmall]{acmart}
\usepackage{mathcomp}

\AtBeginDocument{%
  \providecommand\BibTeX{{%
    \normalfont B\kern-0.5em{\scshape i\kern-0.25em b}\kern-0.8em\TeX}}}

\setcopyright{none}
\copyrightyear{2020}
\acmYear{2020}
\acmDOI{}

\begin{document}

\title{Uncovering socioeconomic gaps in mobility reduction during the COVID-19 pandemic using location data}

\author{Samuel P. Fraiberger}
\email{sfraiberger@worldbank.org}
\affiliation{%
  \institution{World Bank, Massachusetts Institute of Technology, New York University}
}

\author{Pablo Astudillo}
\affiliation{%
  \institution{World Bank, Oxford University}
}
\email{pastudilloesteve@worldbank.org}

\author{Lorenzo Candeago}
\affiliation{%
 \institution{World Bank, Fondazione Bruno Kessler}
}
\email{lcandeago@worldbank.org}

\author{Alex Chunet}
\affiliation{%
  \institution{World Bank}
}
\email{achunet@worldbank.org}

\author{Nicholas K. W. Jones}
\affiliation{%
  \institution{World Bank}
  }
\email{njones@worldbankgroup.org}

\author{Maham Faisal Khan}
\affiliation{%
  \institution{World Bank}
  }
\email{mkhan57@worldbank.org}

\author{Bruno Lepri}
\affiliation{%
 \institution{Fondazione Bruno Kessler}
}
\email{lepri@fbk.eu}

\author{Nancy Lozano Gracia}
\affiliation{\institution{World Bank}}
\email{nlozano@worldbank.org}

\author{Lorenzo Lucchini}
\affiliation{%
 \institution{Fondazione Bruno Kessler}
}
\email{llucchini@fbk.eu}

\author{Emanuele Massaro}
\affiliation{\institution{World Bank, Massachusetts Institute of Technology, \'Ecole Polytechnique F\'ed\'erale de Lausanne}}
\email{emassaro@worldbank.org}

\author{Aleister Montfort}
\affiliation{\institution{World Bank}}
\email{amontfortibieta@worldbank.org}

\renewcommand{\shortauthors}{Fraiberger, et al.}

\begin{abstract}
  Using smartphone location data from Colombia, Mexico, and Indonesia, we investigate how non-pharmaceutical policy interventions intended to mitigate the spread of the COVID-19 pandemic impact human mobility. In all three countries, we find that following the implementation of mobility restriction measures, human movement decreased substantially. Importantly, we also uncover large and persistent differences in mobility reduction between wealth groups: on average, users in the top decile of wealth reduced their mobility up to twice as much as users in the bottom decile. For decision-makers seeking to efficiently allocate resources to response efforts, these findings highlight that smartphone location data can be leveraged to tailor policies to the needs of specific socioeconomic groups, especially the most vulnerable.
\end{abstract}

\ccsdesc[500]{Information systems~Location based services}
\ccsdesc[500]{Social and professional topics~Economic impact}

\keywords{COVID-19, mobility, physical distancing, GPS location data}


\maketitle

\section{Introduction}
2020 has seen the SARS-CoV-2 infect millions of people and cause unprecedented disruptions around the world, pushing healthcare services to their limits and grinding economies to a halt. Pursuant to guidelines issued by the World Health Organization (WHO), national and local governments worldwide have implemented social distancing policies ranging from strict national lockdowns to limited curfew hours, in an attempt to curb the virus reproduction rate and prevent the overburdening of health care systems. It has been shown that these policies---\textit{non-pharmaceutical interventions (NPIs)}---can help reduce the transmission of the virus by limiting contacts between individuals \cite{chinazzi2020travel, flaxman2020npi, kraemer2020effect, lai2020effect, salje2548181estimating, zhang2020contactpatterns}.

In this paper, we use GPS location data to characterize changes in human mobility as the pandemic unfolded. Researchers and policy analysts are now routinely using GPS location data collected from smartphones to quantify human mobility patterns in real-time and with high spatial precision. Despite comparatively less coverage in low- and middle-income countries, these data sources are nonetheless of immense value to improve our understanding of how the current crisis is unraveling \cite{oliver2020mobile}. 

We focus our work on three developing countries in two different continents: Colombia, Mexico, and Indonesia. Our main contribution is to combine spatio-temporal trajectories of smartphone users with socio-economic information at high spatial granularity, allowing us to characterize differences in mobility patterns across wealth groups. Taken together, our findings show large and persistent differences in mobility reduction between wealth groups: on average, users in the top decile of wealth reduced their mobility up to twice as much as users in the bottom decile as a result of NPIs.

In the reminder of the paper, we start by contextualizing our work within the broader efforts by the research community to quantify mobility changes during the COVID-19 pandemic. We then provide an overview of data and methods used, followed by a description of our main results. We close by discussing the relevance of this work for policy-making and decision support in developing countries. 

\section{Background}
Since the onset of the COVID-19 pandemic, governments, both national and local, have enacted social distancing measures to varying levels of stringency \cite{hale2020variation}. 
To quantify changes in mobility owing to these policy measures, several large-scale technology companies have made available anonymized location data to support pandemic response and recovery. Google's `Community Mobility Maps'\footnote{\url{https://www.google.com/covid19/mobility/}}, Facebook's expansion of its previous `Disaster Maps' offerings\footnote{\url{https://about.fb.com/news/2020/04/data-for-good/}}\cite{maas2019facebook}, Apple's `Mobility Trends Reports'\footnote{\url{https://www.apple.com/covid19/mobility}}, as well as data pipelines provided by TomTom\footnote{\url{https://www.tomtom.com/covid-19/}}, Unacast\footnote{\url{https://www.unacast.com/covid19}} and Cuebiq\footnote{https://www.cuebiq.com/visitation-insights-covid19/}, are only a few examples of these efforts.

A global community of researchers  has quickly mobilized to engage these available datasets to quantify the impact of social distancing policies and better understand their influence on the spread of the virus \cite{buckee2020aggregated, oliver2020mobile}. Abouk and Heydari \cite{abouk2020immediate} outlined the ways in which different types of location-based mobility information can help to better target and design measures to contain and slow the spread of the COVID-19 pandemic. Together with daily state-level data on COVID-19 tests and confirmed cases, they rank policies based on their effectiveness using a difference-in-differences approach. They found that statewide stay-at-home orders had the strongest causal impact on reducing social interactions. Klein \emph{et al.} \cite{klein2020assessing} analyzed mobility data for the US and find that, by March 23, the policies enacted had reduced by half the overall mobility in several major US cities. Along the same line, Pepe \emph{et al.} \cite{pepe2020covid} analyzed GPS data in Italy and found that the restrictions in mobility, closure of public spaces, and the enhancement of smart/remote working, led to an average reduction of potential encounters of 8\% during week the second week and almost 19\% during third week following the outbreak. Kraemer \emph{et al.} \cite{kraemer2020effect} used real-time mobility data from Wuhan along with travel history to investigate case transmission in cities across China and ascertain the impact of implemented policy. Tian \emph{et al.} \cite{tian2020investigation} investigated the spread and control of COVID-19 using a dataset that included case reports, human movement, and public health interventions, and found that the Wuhan shutdown effectively delayed the arrival of COVID-19 in other cities by 2.91 days. 
Maloney and Taskin \cite{MaloneyTaskin2020} used Google mobility data for a large sample of countries across various income levels to assess the determinants of mobility changes during the pandemic. To our knowledge, our paper is the first to quantify the differences in mobility reduction between socioeconomic groups in developing countries. 



\section{DATA AND METHODS}
\subsection{GPS location data}

Anonymized GPS location data for this study was provided by Cuebiq\footnote{https://www.cuebiq.com/}, a location intelligence company. Cuebiq integrates a software development kit into over 86 mobile apps, allowing them to collect data at high precision from users who have opted-in through a GDPR compliant framework. Our dataset contains privacy-preserving, timestamped geolocation data from January 1 to May 7, 2020, from over 270 thousand users in Colombia, 1.38 million users in Mexico, and 276 thousand users in Indonesia, representing about $0.6\%$, $1.1\%$, and $0.1\%$ of the population respectively. Owing to the spatial distribution of cellphone ownership in these countries, about $30\%$ of users are located in their capital cities. 

To quantify the mobility patterns of a homogeneous set of people over time, we focus on users who were active on at least half of the days since January 1, 2020; a user is considered active on any given day if her GPS location was observed at least once. 

\subsection{Mobility indicators}

To preserve user privacy, Cuebiq anonymizes users and randomizes their home location point geometries within a geohash grid cell, thereby allowing demographic analysis based on residential areas without revealing the precise location or identity of these users. We are therefore able to conduct an analysis of time spent at home at the level of the smallest administrative units provided by the most recent population census available in each country\footnote{We use the most recent year available for each country: 2010 for Mexico and Indonesia, and 2018 for Colombia.}. We define a user's home location to be her most frequently visited administrative unit during nighttime hours (11pm-7am) in our period of analysis. We characterize a user's time spent at home to be the duration between consecutive observations within the administrative unit identified as her home location, only considering pairs of observations that are separated by less than 12 hours. 

Next, we define a user's work location to be the most frequently visited administrative unit for each user during daytime hours (9am-6pm) on weekdays. To identify commuters, we compute the share of users who are visiting the administrative unit of their work location. To quantify the number of neighborhood visits of each user, we compute the number of unique administrative units that she visits relative to her number of observed locations. This allows us to capture the diversity of places visited, controlling for the level of activity of each user. We then normalize our measure by the average number of neighborhood visits across users during the period of analysis, for ease of interpretation. Finally, to compute the maximum distance between observed locations, we discard observations for which the GPS accuracy is less than 200 meters. For the estimation of the average value of each mobility indicator on a given day, we include only users who have more than 10 GPS location points on that day.

\subsection{Socioeconomic characterization of users}

To quantify differences in variations in mobility between socioeconomic groups, we construct a wealth index using population census data for each country. First, we compute average values for asset ownership (e.g. car, fridge, computer, etc.) and access to services (e.g. health, education, etc.) within each administrative unit\footnote{The variables used to create the wealth index vary slightly between countries depending on data availability.}. Then, we reduce the data to a one-dimensional wealth index by taking its principal component \cite{vyass2006constructing}. Finally, we proxy each user's wealth using the index value of the administrative unit of her home location.

\section{RESULTS} 
\subsection{Changes in mobility patterns over time}

\begin{figure*}[h]
  \centering
  \includegraphics[width=\textwidth]{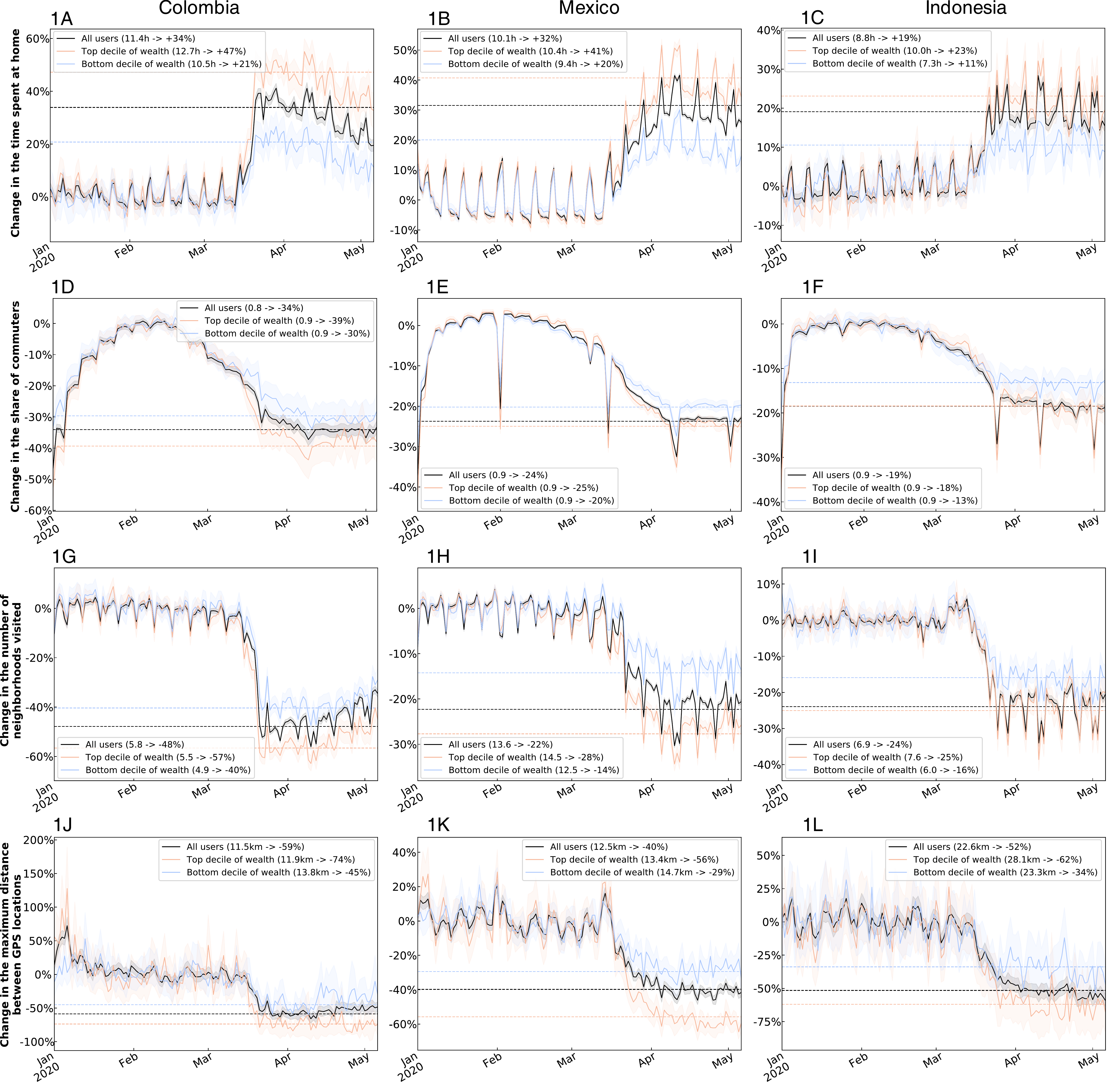}
  \caption{Variations in users' mobility patterns in Colombia (left column), Mexico (middle column), and Indonesia (right column). We computed 4 indicators of mobility: \textbf{A-C)} the average time users spend at home, \textbf{D-F)} the share of users commuting from home to work, \textbf{G-I)} the average number of neighborhoods (defined by administrative units) in which they are active, and \textbf{J-L)} the maximum distance between their GPS locations. Users are divided into deciles based on the value of the wealth index of the admin unit where they live. For each mobility indicator, we report changes in mobility relative to the average mobility computed during the first 2 weeks of February for all users (black), users in the top decile (orange) and users in the bottom decile (blue). Average value during the first 2 weeks of February, and the change relative to the first 2 weeks of April are reported in parentheses and indicated by the horizontal dotted lines. Taken together, this figure documents a large and persistent mobility gap between users based on their wealth.}
  
  \Description{}
  \label{fig1}
\end{figure*}

In this section we quantify the extent to which user movements decreased since the onset of the COVID-19 pandemic. Fig \ref{fig1}A, B and C indicate sharp increases in time spent at home: in Colombia from March 17 when a state of emergency was declared; in Mexico from the earliest introductions of social distancing policies in mid-March; and in Indonesia from March 15 when the President issued a nationwide advisory to stay at home. Between the second week of February and the second week of April, the time users spent at home increased by 34\%, 32\%, and 19\% in Colombia, Mexico, and Indonesia respectively. In all three countries, sharp differences are visible between the top and bottom deciles of wealth ("high-wealth users"): the rate of change for users in the top decile is about twice as large as that of users in the bottom decile ("low-wealth users").

In addition to spending more time at home, a large share of users stopped commuting to work on a daily basis [Fig. \ref{fig1}D-F]: the share of commuters decreased between mid-February and mid-April by 34\%, 24\% and 19\% in Colombia, Mexico, and Indonesia respectively. Whilst the mobility gap between high- and low-wealth users is smaller when looking at commuting patterns than when focusing on the time users spend at home, it is nonetheless substantial: the rate of reduction in commuting is 25 to 30\% greater for high-wealth users than for those on the lower end of the wealth distribution. 

We also found that the average number of neighborhoods that a user visits dropped sharply between mid-February and mid-April [Fig. \ref{fig1}G-H]. In Indonesia, we initially see close to a 5\% increase in the number of neighborhoods visited for both the top and bottom deciles of wealth, potentially indicating that users anticipating a possible lockdown decided to increase their movement. Immediately after this, the number of neighborhoods visited decreased and stabilized in a manner similar to Mexico. In Colombia, mirroring the trend observed in the time users spend at home, the number of neighborhoods visited started increasing again after the sharp decrease that happened in mid-March.\footnote{Other national observances are also visible in the form of sharp drops in commuter data, including Constitution Day on February 03, Benito Juarez's Birthday on March 16, Maundy Thursday/Good Friday on Apr 09-10, and Labor Day on May 01. Good Friday on April 10 and Waisak Day on May 07 are also low commuter days for Indonesia.}
Overall, the number of neighborhoods visited decreased between 50 and 100\%  more for high-wealth users compared to low-wealth users. 

Finally, the maximum distance between a user's GPS locations on a typical day decreased by 59\%, 40\% and 52\% in Colombia, Mexico, and Indonesia respectively [Fig. \ref{fig1}I-L]. 
Consistent with previous results, we found that high-wealth users reduced their traveled distance almost twice as much as low-wealth users.

\section{Discussion}

Our findings highlight that, even in developing countries where smartphone penetration is more limited compared to what is observed in the developed world, highly precise location data can shed light on changes in mobility patterns, providing an instrument for policymakers aiming to efficiently allocate resources to response mechanisms. Our findings also suggest that, as a measure to protect health and livelihoods, lockdown measures may be excessively costly for the most vulnerable, particularly in the developing world, where large proportions of the population live hand-to-mouth, crowding levels are high, and the availability of services limited. 



In light of these complex challenges, complementary policies are needed to help protect the most vulnerable. The analysis presented here can provide information about the groups and areas that may need targeted interventions, and where the provision of additional services complementing social distancing measures is warranted. Such services could include dedicated information campaigns to increase awareness, and additional resources assigned to protect those that cannot stay home. By looking at the mobility patterns of those that are constrained to move around the city during a lockdown, policymakers could think about how to safely provide bus routes services to maintain connectivity between areas that are seeing sustained flows. In overcrowded areas where staying home is difficult, policymakers could think about investments in open spaces, and through that facilitate the implementation of social distancing measures. This type of analysis could also be used as input to assess the risks and immediate impacts of lifting lockdown measures. Thinking ahead, policymakers can use the analysis in this paper to identify areas that may require additional investments to build long-term resilience through better services. 

\begin{acks}
We extend our sincere gratitude to Cuebiq for providing the data to support this effort. This work has been supported by the World Bank Group Partnership Fund for the SDGs. The findings, interpretations, and conclusions expressed in this paper are entirely those of the authors. They do not necessarily represent the views of the International Bank for Reconstruction and Development/World Bank and its affiliated organizations, or those of the Executive Directors of the World Bank or the governments they represent.
\end{acks}

\bibliographystyle{ieeetr}
\bibliography{sample-base}

\end{document}